\documentclass[prd,reprint,nofootinbib,superscriptaddress,longbibliography]{revtex4-2}
\usepackage{bm}
\usepackage{amsmath}
\usepackage{amssymb}
\usepackage{graphicx}
\usepackage{subfigure}
\usepackage{hyperref}
\usepackage{multirow}
\usepackage{color}
\usepackage{booktabs}
\usepackage{amsmath}
\usepackage{enumitem} 
\usepackage{siunitx}
\usepackage{CJK}
\usepackage{orcidlink}
\usepackage{array}
\usepackage{indentfirst}
\usepackage{float}

\hypersetup{
	colorlinks=true,
	linkcolor=red,
	citecolor=blue,
}

\begin{document}

\begin{CJK*}{UTF8}{gbsn}

\title{Frequency-domain extended-effective-source gravitational self-force for eccentric Schwarzschild orbits}
\author{Xuchen Lu (路旭晨) 
\orcidlink{0000-0002-9093-9059}}
\email{luxuchen@nbu.edu.cn}
\affiliation{Zhejiang Key Laboratory of Extreme Universe,
Institute of Fundamental Physics and Quantum Technology,
Department of Physics, School of Physical Science and Technology, Ningbo
University, Ningbo, Zhejiang 315211, China}

\author{Yungui Gong (龚云贵)
\orcidlink{0000-0001-5065-2259}}
\email{gongyungui@nbu.edu.cn}
\affiliation{Zhejiang Key Laboratory of Extreme Universe,
Institute of Fundamental Physics and Quantum Technology, Department of Physics, School of Physical Science and Technology, Ningbo University, Ningbo, Zhejiang 315211, China}

\author{Bokai Zhang (张博凯)
\orcidlink{0009-0004-8123-1894}}
\email{bkzhang07@163.com}
\affiliation{School of Physics, Huazhong University of Science and Technology, 1037 LuoYu Rd, Wuhan, Hubei 430074, China}

\author{Chao Zhang (张超)
\orcidlink{0000-0001-8829-1591}}
\email{zhangchao1@nbu.edu.cn}
\affiliation{Zhejiang Key Laboratory of Extreme Universe,
Institute of Fundamental Physics and Quantum Technology, Department of Physics, School of Physical Science and Technology, Ningbo University, Ningbo, Zhejiang 315211, China}

\author{Wenting Zhou (周文婷)
\orcidlink{0000-0003-4046-753X}}
\email{zhouwenting@nbu.edu.cn}
\affiliation{Zhejiang Key Laboratory of Extreme Universe,
Institute of Fundamental Physics and Quantum Technology, Department of Physics, School of Physical Science and Technology, Ningbo University, Ningbo, Zhejiang 315211, China}

\begin{abstract}
Effective-source formulations provide a practical route to gravitational self-force (GSF) calculations when singular retarded fields cannot be handled directly mode by mode.
Before this work, frequency-domain effective-source calculations of the first-order GSF had been performed only for circular Schwarzschild orbits.
We construct and implement a frequency-domain extended effective-source (EES) formulation for first-order Lorenz-gauge GSF on eccentric orbits in Schwarzschild spacetime.
The central obstacle is that radial libration forces the physical puncture and effective source to switch between interior and exterior branches, limiting their differentiability and producing slow spectral convergence.
We overcome this obstruction by analytically extending both branches across the libration region and solving the coupled Lorenz-gauge perturbation equations for the resulting smooth extended sources.
This provides an end-to-end gravitational EES implementation for eccentric motion and validates a computational ingredient needed for extending frequency-domain effective-source calculations toward second order.
\end{abstract}

\maketitle

\section{Introduction}

Extreme-mass-ratio inspirals (EMRIs), in which a stellar-mass compact object spirals into a massive black hole, are among the most important sources for future space-based gravitational-wave observatories, including LISA \cite{Danzmann:1997hm,LISA:2017pwj}, TianQin \cite{TianQin:2015yph}, and Taiji \cite{Hu:2017mde}.
Because an EMRI can execute tens of thousands of relativistic orbital cycles in the strong-field region before plunge, its gravitational waveform carries
detailed information about the spacetime in the immediate vicinity of the central object.
EMRI observations can therefore provide precision probes of the central black-hole geometry and of possible environmental or nonvacuum effects, while simultaneously enabling stringent tests of strong-field gravity \cite{Babak:2017tow,Berry:2019wgg,Destounis:2020kss,Cardoso:2021wlq,Dai:2023cft,Zhang:2024ogc,Zhao:2024bpp,Zhao:2026yis}.

Realizing this science requires correspondingly accurate models of the orbital evolution and gravitational waveform.  
The extreme mass ratio provides a natural perturbative parameter, $\epsilon=\mu/M\sim 10^{-7}-10^{-4}$, where $\mu$ and $M$ are the masses of the small compact object and the central supermassive black hole, respectively.  
At leading order the small body follows a geodesic of the background spacetime.
At subsequent orders its own gravitational field perturbs that spacetime and acts back on its motion through the gravitational self-force (GSF).  
GSF theory provides a systematic framework for calculating this post-geodesic dynamics \cite{Mino:1996nk,Quinn:1996am,Detweiler:2002mi,
Poisson:2011nh,Barack:2018yvs,Pound:2021qin,LISAConsortiumWaveformWorkingGroup:2023arg}.
First-order GSF calculations have now reached a high level of maturity,
including calculations for circular orbits \cite{Barack:2007tm,Detweiler:2008ft,Akcay:2010dx}, eccentric orbits in Schwarzschild spacetime \cite{Warburton:2011fk,Akcay:2013wfa,Osburn:2014hoa,Osburn:2015duj,Warburton:2017sxk},
and generic bound motion in Kerr spacetime \cite{vandeMeent:2017bcc,VanDeMeent:2018cgn,Hinderer:2008dm,Miller:2020bft,Fujita:2020zxe,Lynch:2021ogr,Hughes:2021exa,Lynch:2023gpu}.
Self-consistent evolution has been studied theoretically for gravitational self-force \cite{Pound:2009sm},
but so far implemented numerically only in scalar-field toy models \cite{Diener:2011cc,Wittek:2024gxn}.

A central difficulty in self-force theory is that the retarded field of a point particle diverges on the particle's worldline, 
precisely where the force must be evaluated.  
The practical problem is therefore one of regularization:
the singular contribution must be isolated and removed with sufficient accuracy to recover the regular field.
At first perturbative order, one of the most successful implementations of this regularization is the mode-sum method \cite{Barack:1999wf,Barack:2001bw}.
This method expands the full retarded field into spherical-harmonic modes, analytically characterizes the singular contribution to each mode via regularization parameters, and subtracts it mode by mode, yielding a finite mode sum.

First-order accuracy, however, is not sufficient for the full precision requirements of EMRI waveform modeling.
The number of orbital cycles accumulated during an EMRI scales with the inverse mass ratio, $\epsilon^{-1}=M/\mu$.
Two-timescale analyses further show that, over the radiation-reaction timescale, the leading adiabatic term contributes $O(\epsilon^{-1})$ to the waveform phase, while the first-post-adiabatic term contributes $O(\epsilon^{0})$ radians \cite{Hinderer:2008dm,Miller:2020bft}.
This $O(\epsilon^{0})$ phase contribution depends in part on the dissipative component of the second-order GSF, motivating the development of second-order GSF theory and methods \cite{Rosenthal:2006nh,Rosenthal:2006iy,Pound:2012nt,Pound:2019lzj,Wardell:2021fyy,Upton:2021oxf,Detweiler:2011tt,Gralla:2012db,Pound:2012dk,Pound:2015wva,Pound:2017psq,Lousto:2008vw,Wardell:2015kea,Upton:2023tcv,Spiers:2023mor,Spiers:2023cip,Miller:2023ers,Leather:2026zhl,Wei:2025lva,Pound:2014xva,Miller:2016hjv}.
A fundamental complication is that the convenient regularity property underlying first-order mode-sum regularization does not generally survive in conventional second-order formulations: 
individual multipoles of the second-order retarded field can themselves remain singular on the worldline \cite{Pound:2014xva},
while the quadratic source introduces the additional complication of infinite mode coupling. 
Consequently, the standard first-order mode-sum prescription does not directly provide a practical second-order regularization scheme.

The effective-source approach provides a natural alternative \cite{Vega:2007mc,Barack:2007jh,Leather:2023dzj}.
Rather than first solving for a singular retarded field and regularizing it only at the particle's location, one constructs a local puncture field approximating the Detweiler--Whiting singular field \cite{Detweiler:2002mi} and defines a residual field satisfying a wave equation whose effective source is finite, with regularity determined by the order of the puncture. 
The calculation therefore solves directly for a regularized field and avoids the direct numerical treatment of the singular point-particle source.
This feature makes effective-source formulations particularly attractive for extending GSF calculations to perturbative orders at which a harmonic decomposition alone does not sufficiently regularize the retarded field.
Warburton and Wardell incorporated effective-source regularization into a frequency-domain calculation for a scalar charge on a circular Schwarzschild orbit \cite{Warburton:2013lea},
and subsequently extended the approach to Lorenz-gauge gravitational perturbations and the first-order GSF for circular Schwarzschild orbits \cite{Wardell:2015ada}.  
These calculations established the viability of the frequency-domain effective-source approach for gravitational self-force calculations.

Eccentricity is astrophysically important: EMRIs formed through stellar dynamical capture can retain appreciable eccentricity in the observable band \cite{Barack:2003fp,Hopman:2005vr,Amaro-Seoane:2007osp},
and eccentricity can produce measurable structure in their gravitational waveforms \cite{Destounis:2021mqv,Destounis:2021rko}.
For eccentric orbits, radial libration introduces an infinite set of harmonics with frequencies $\omega_{mn}=m\Omega_\phi+n\Omega_r$.
At any fixed radius inside the libration region, the particle crosses that radius twice per radial period.
The physical puncture and effective source therefore switch between interior and exterior branches as the particle passes through that radius.
This switching limits their differentiability as functions of time, causing algebraically decaying Fourier coefficients and slow convergence, together with Gibbs-type behavior near the worldline. 
Related convergence difficulties for distributional point-particle sources can be overcome with the method of extended homogeneous solutions \cite{Barack:2008ms,Akcay:2013wfa},
but that construction cannot be transferred directly to an effective-source worldtube containing an inhomogeneous particular solution.

Leather and Warburton overcame this obstruction for a scalar charge on an eccentric Schwarzschild orbit by introducing the extended effective-source (EES) method \cite{Leather:2023dzj}.
Instead of Fourier decomposing the finitely differentiable physical source directly, the interior and exterior source and puncture expressions are analytically extended across the libration region.
The resulting extended quantities are smooth and possess rapidly convergent spectral representations;
the physical residual field is recovered only after the two extended solutions have been reconstructed.
For the gravitational problem, a crucial additional ingredient was recently provided by Zhang \textit{et al.} \cite{Zhang:2025eqz}, who constructed an analytic first-order Lorenz-gauge gravitational puncture and effective source for generic geodesic motion in Schwarzschild spacetime.
That construction provides explicit expressions for all ten tensor-harmonic components and was subsequently validated in a time-domain effective-source calculation of the GSF for circular Schwarzschild orbits \cite{Zhang:2026gdk}.
For frequency-domain self-force calculations, however, these time-domain expressions must be Fourier decomposed.

In this work, we combine these developments to construct and implement a frequency-domain gravitational EES formulation for bound eccentric orbits in Schwarzschild spacetime. 
Building on the generic-orbit gravitational puncture and effective source of Ref. \cite{Zhang:2025eqz} and the scalar EES strategy of Ref. \cite{Leather:2023dzj},
we analytically extend the two gravitational source branches across the radial libration region,
solve the resulting smooth frequency-domain sources within the coupled Lorenz-gauge system, reconstruct the physical residual metric perturbation, and evaluate the first-order GSF directly from that regularized field.

We emphasize that the first-order eccentric Schwarzschild GSF was already obtained with the mode-sum method and is used here as an independent benchmark. 
The advance demonstrated here is the first end-to-end frequency-domain gravitational EES implementation for eccentric motion.
It removes the spectral obstruction caused by radial branch switching and provides a concrete validation of the regularized-field architecture needed to extend frequency-domain effective-source calculations toward second perturbative order.

The remainder of this paper is organized as follows.
In Sec. \ref{sec.Orbital}, we review bound eccentric geodesics in Schwarzschild spacetime and the Lorenz-gauge tensor-harmonic field equations. In Sec. \ref{sec.efs}, we introduce the gravitational puncture and effective source and describe their Fourier and Chebyshev representations.
In Sec. \ref{sec:EES}, we develop the extended effective-source construction and the reconstruction of the physical residual field. 
Section \ref{sec.nir} presents the numerical implementation, self-force evaluation, and validation results.
The conclusion is in Sec. \ref{sec:conclusion}.

\section{Orbital Dynamics and Lorenz-Gauge Perturbations}
\label{sec.Orbital}

\subsection{Bound eccentric geodesics in Schwarzschild spacetime}

We consider a non-spinning compact object of mass $\mu$ orbiting a Schwarzschild black hole of mass $M$, 
with mass ratio $\epsilon\equiv\mu/M\ll1$. 
Throughout this work, we set $G=c=M=1$ for convenience.

In Schwarzschild coordinates $x^\mu=(t,r,\theta,\phi)$, 
the background line element is
\begin{equation}
ds^2=-f(r)dt^2+f(r)^{-1}dr^2+r^2\left(d\theta^2+\sin^2\theta\,d\phi^2\right),
\end{equation}
where $f(r)=1-2/r$.
The particle worldline $x_p^\mu(\tau)$ is parametrized by proper time $\tau$,  and its four-velocity $u^\mu=dx_p^\mu/d\tau$ satisfies the normalization condition $u^\mu u_\mu=-1$.
At zeroth order in the mass ratio, the small body follows a bound timelike geodesic of the Schwarzschild background. 
Without loss of generality, we restrict the motion to the equatorial plane,
$\theta_p=\pi/2$.
The conserved specific energy and angular momentum are
$\mathcal{E}=-u_t$ and $\mathcal{L}=u_\phi$.

We parametrize the orbit by semilatus rectum $p$, eccentricity $e$ and relativistic anomaly $\chi$,
\begin{equation}
r_p(\chi)=\frac{p}{1+e\cos\chi}.
\end{equation}
The particle librates between
$r_{\min}=p/(1+e)$ and $r_{\max}=p/(1-e)$.
Stable bound geodesics satisfy
$0\leq e<1$ and $p>6+2e$.
The coordinate-time and azimuthal motions are determined by
\begin{equation}
\begin{split}
\frac{dt_p}{d\chi}=&\frac{p^2}{(1+e\cos\chi)^2(p-2-2e\cos\chi)}\\
&\times \sqrt{\frac{(p-2)^2-4e^2}{p-6-2e\cos\chi}},    
\end{split}
\end{equation}
and
\begin{equation}
\frac{d\phi_p}{d\chi}=\sqrt{\frac{p}{p-6-2e\cos\chi}}.
\end{equation}
The radial period in Schwarzschild coordinate time and the accumulated azimuthal angle over one radial libration are
\begin{equation}
T_r=\int_0^{2\pi}\frac{dt_p}{d\chi}\,d\chi,\quad\Phi=\int_0^{2\pi}\frac{d\phi_p}{d\chi}\,d\chi.
\end{equation}
The corresponding fundamental frequencies are
\begin{equation}
\Omega_r=\frac{2\pi}{T_r},
\quad
\Omega_\phi=\frac{\Phi}{T_r}.
\end{equation}
They determine the discrete frequency spectrum
\begin{equation}
\label{eq:frequency_spectrum}
\omega_{mn}=m\Omega_\phi+n\Omega_r,
\quad
m,n\in\mathbb Z.
\end{equation}

\subsection{Tensor-harmonic field equations}
We describe the first-order metric perturbation using its trace-reversed form
\begin{equation}
\bar h_{\mu\nu} = h_{\mu\nu}-\frac{1}{2}g_{\mu\nu}h,
\quad
h=g^{\rho\sigma}h_{\rho\sigma}.
\end{equation}
In the Lorenz gauge, the perturbation satisfies
$\nabla^\mu\bar h_{\mu\nu}=0$,
together with the linearized Einstein equation
\begin{equation}
\label{eq:linearized_einstein}
\Box\bar h_{\mu\nu}+2R_{\mu\rho\nu\sigma}\bar h^{\rho\sigma}=-16\pi T_{\mu\nu}.
\end{equation}
Here, $\Box=\nabla^\rho\nabla_\rho$ and
$R_{\mu\rho\nu\sigma}$ is the Riemann tensor of the Schwarzschild background. 
The point-particle stress-energy tensor is
\begin{equation}
T^{\mu\nu}(x)=\mu\int\frac{u^\mu u^\nu}{\sqrt{-g}}\,
\delta^{(4)}\!\left[x-x_p(\tau)\right]d\tau .
\end{equation}

Following Refs. \cite{Barack:2005nr,Barack:2007tm}, we decompose the trace-reversed perturbation in the Barack--Lousto--Sago tensor-harmonic basis $Y_{\mu\nu}^{(i)\ell m}$,
which incorporates the redefinition of the $i=3$ field, 
and use the coupling operators of the gauge-damped field equations, 
\begin{equation}
\bar h_{\mu\nu}=\frac{\mu}{r}\sum_{\ell=0}^{\infty}\sum_{m=-\ell}^{\ell}\sum_{i=1}^{10}a_\ell^{(i)}\bar h_{\ell m}^{(i)}(t,r)Y_{\mu\nu}^{(i)\ell m}(\theta,\phi;r),
\label{eq:supp_tensor_decomposition}
\end{equation}
where the normalization coefficients are
\begin{equation}
a_\ell^{(i)}
=
\begin{cases}
1/\sqrt{2},
&
i=1,2,3,6,
\\
1/\sqrt{2\ell(\ell+1)},
&
i=4,5,8,9,
\\
1/\sqrt{2(\ell-1)\ell(\ell+1)(\ell+2)},
&
i=7,10.
\end{cases}
\end{equation}
The stress-energy tensor is decomposed in the same basis,
\begin{equation}
T_{\mu\nu}=
\sum_{\ell m}
\sum_{i=1}^{10}
T^{(i)}_{\ell m}(t,r)Y_{\mu\nu}^{(i)\ell m}(\theta,\phi;r).
\end{equation}

Projecting Eq. \eqref{eq:linearized_einstein} onto this basis gives ten coupled $(1+1)$-dimensional equations,
\begin{equation}
\label{eq:box2d}
\square_{2d}\bar h^{(i)}_{\ell m}+
\mathcal{M}^{(i)}_{\ (j)}\bar h^{(j)}_{\ell m}=
4\pi\mu^{-1}\left(\frac{rf}{a^{(i)}_\ell}\right)
T^{(i)}_{\ell m}\equiv S^{(i)}_{\ell m},
\end{equation}
where $i=1,\ldots,10$.
The two-dimensional wave operator $\square_{2d}$ is
\begin{equation}
    \begin{split}
\Box_{2d}=\frac{1}{4}
\left[\frac{\partial^2}{\partial t^2}-f f' \frac{\partial}{\partial r}-f^2 \frac{\partial^2}{\partial r^2}+V_{\ell}(r)\right],\\
V_\ell(r)=f(r)\left[\frac{2}{r^3}+\frac{\ell(\ell+1)}{r^2}\right],
  \end{split}
\end{equation}
where $f'= df/dr$, and $\mathcal M^{(i)}{}_{(j)}$ contains the couplings among the tensor-harmonic components.
For example, the $(i=3)$ coupling term is
\begin{equation}
\begin{split}
\mathcal{M}_{(j)}^{(3)}\bar{h}_{\ell m}^{(j)}
= -\frac{f}{2r^{2}}\Bigl[
&\bar{h}_{\ell m}^{(1)}-\bar{h}_{\ell m}^{(5)}
\\
&-\left(1-\frac{4}{r}\right)
 \left(\bar{h}_{\ell m}^{(3)}+\bar{h}_{\ell m}^{(6)}\right)
\Bigr].
\end{split}
\end{equation}
The remaining coupling operators are given in Appendix A of Ref. \cite{Barack:2007tm}.
The system separates into even- and odd-parity sectors. The components $i=1,\ldots,7$ belong to the even-parity sector, while $i=8,9,10$ belong to the odd-parity sector.
The even-parity sources vanish when $\ell+m$ is odd, and the odd-parity sources vanish when $\ell+m$ is even.

\subsection{Frequency-domain decomposition and hierarchical reduction}

Using the discrete spectrum \eqref{eq:frequency_spectrum}, we write
\begin{equation}
\begin{split}
\bar h_{\ell m}^{(i)}(t,r)
&=
\sum_{n=-\infty}^{\infty}
R_{\ell mn}^{(i)}(r)e^{-i\omega_{mn}t},
\\
S_{\ell m}^{(i)}(t,r)
&=
\sum_{n=-\infty}^{\infty}
S_{\ell mn}^{(i)}(r)e^{-i\omega_{mn}t}.
\end{split}
\end{equation}
For each $(\ell,m,n)$ mode, the field equations reduce schematically to
\begin{equation}
\begin{split}
\left[\frac{d^2}{dr_*^2}+\omega_{mn}^2-V_\ell(r)\right]
R_{\ell mn}^{(i)}
&-4\sum_j \mathcal{M}^{(i)}{}_{(j)}(r)R_{\ell mn}^{(j)}
\\
&=-4 S_{\ell mn}^{(i)}(r),
\end{split}
\end{equation}
where the tortoise coordinate is
\begin{equation}
\frac{dr_*}{dr}=f^{-1}(r),
\quad
r_*=r+2\ln \left(\frac{r}{2}-1\right),
\end{equation}
and the frequency-domain coupling operators $\mathcal M^{(i)}{}_{(j)}$ are given in Appendix A of Ref. \cite{Akcay:2013wfa}.

The Lorenz-gauge condition provides four constraint equations that allow a subset of the radial functions to be reconstructed algebraically. 
For simplicity, we suppress the mode indices $(\ell,m,n)$ on the radial functions and write $\omega\equiv\omega_{mn}$.
These constraints are
\begin{equation}
\begin{split}
&i\omega R^{(1)}
 +f\Bigl(i\omega R^{(3)}+R^{(2)}_{,r}
 +\frac{R^{(2)}-R^{(4)}}{r}\Bigr)=0,\\
&-i\omega R^{(2)}-fR^{(1)}_{,r}+f^2R^{(3)}_{,r}\\
&\qquad
 -\frac{f}{r}\Bigl(R^{(1)}-R^{(5)}
 -fR^{(3)}-2fR^{(6)}\Bigr)=0,\\
&-i\omega R^{(4)}
 -\frac{f}{r}\Bigl(rR^{(5)}_{,r}+2R^{(5)}
 +\ell(\ell+1)R^{(6)}-R^{(7)}\Bigr)=0,\\
&-i\omega R^{(8)}
 -\frac{f}{r}\Bigl(rR^{(9)}_{,r}
 +2R^{(9)}-R^{(10)}\Bigr)=0.
\end{split}
\end{equation}

We exploit the hierarchical structure of the coupled Lorenz-gauge equations to construct the homogeneous solutions \cite{Akcay:2013wfa}.
For generic dynamical modes with $\omega_{mn}\neq0$,
we integrate only a minimal numerical set of coupled radial equations and use the Lorenz-gauge constraints to reconstruct the remaining algebraic set.
In the odd-parity sector with $\ell\geq2$, the numerical set is $\{R^{(9)},R^{(10)}\}$, while $R^{(8)}$ is reconstructed algebraically. 
For even-parity modes with $\ell\geq2$, the numerical set is $\left\{
R^{(1)},R^{(3)},R^{(5)},R^{(6)},R^{(7)}
\right\}$,
while $R^{(2)}$ and $R^{(4)}$ are reconstructed from the constraints. 
For dynamical even-parity dipole modes,
the numerical set reduces to $\{R^{(1)},R^{(3)},R^{(5)},R^{(6)}\}$,
and for the dynamical monopole it reduces to $\{R^{(1)},R^{(3)},R^{(6)}\}$.

Static modes require separate treatment because the algebraic reduction used for dynamical modes becomes degenerate when $\omega_{mn}=0$.
For generic nonresonant orbits, the only static mode is $(m,n)=(0,0)$. 
In the odd-parity static sector, $R^{(9)}$ and $R^{(10)}$ vanish, while $R^{(8)}$ is obtained analytically. 
In the even-parity static sector with $\ell\geq2$, the numerical set is $\{R^{(1)},R^{(3)},R^{(5)}\}$, and the remaining functions are reconstructed using the static Lorenz-gauge constraints.
The static monopole is obtained entirely from the known analytic solution. 
The complete hierarchy is summarized in Table \ref{tab:HSS}.

For dynamical modes, the horizon and infinity bases of homogeneous solutions satisfy the retarded boundary conditions
\begin{equation}
R^{(i)}(r)\sim e^{-i\omega r_*},
\quad r_*\rightarrow-\infty,
\end{equation}
and
\begin{equation}
R^{(i)}(r)\sim e^{+i\omega r_*},\quad
r_*\rightarrow+\infty,
\end{equation}
respectively.
The corresponding boundary data are constructed from near-horizon Frobenius expansions and large-radius asymptotic expansions \cite{Akcay:2010dx,Akcay:2013wfa}. 
For static modes, the radiative boundary conditions are replaced by regularity at the future event horizon and spatial infinity.

\begin{table}[H]
\centering
\caption{Hierarchical reduction scheme for the radial functions
$R_{\ell mn}^{(i)}$. Entries in the last two columns list the field
indices $i$: $X\to Y$ means that the fields in $X$ are obtained
directly and used to reconstruct those in $Y$ through the
Lorenz-gauge constraints. The label (A) denotes an analytic solution,
and $\ast$ denotes all $n$.}
\label{tab:HSS}
\small
\setlength{\tabcolsep}{3pt}
\renewcommand{\arraystretch}{1.1}
\begin{tabular*}{\columnwidth}
{@{\extracolsep{\fill}}lcc@{}}
\toprule
$(\ell,m,n)$ & $\ell+m$ even & $\ell+m$ odd \\
\midrule
$(0,0,0)$
  & $1,3\to6$ (A) & $\cdots$ \\
$(0,0,n\ne0)$
  & $1,3,6\to2$ & $\cdots$ \\
\addlinespace
$(1,0,0)$
  & $\cdots$ & $8$ (A) \\
$(1,0,n\ne0)$
  & $\cdots$ & $9\to8$ \\
$(1,\pm1,\ast)$
  & $1,3,5,6\to2,4$ & $\cdots$ \\
\addlinespace
$(\ell\ge2,0,0)$
  & $1,3,5\to6,7$ & $8$ (A) \\
$(\ell\ge2,0,n\ne0)$
  & $1,3,5,6,7\to2,4$ & $9,10\to8$ \\
$(\ell\ge2,m\ne0,\ast)$
  & $1,3,5,6,7\to2,4$ & $9,10\to8$ \\
\bottomrule
\end{tabular*}
\end{table}

\section{Gravitational Effective-Source Regularization}
\label{sec.efs}

\subsection{Self-force}

In first-order perturbation theory, the gravitational self-force is determined by a smooth regular field, as follows from the formulations of Mino, Sasaki, and Tanaka \cite{Mino:1996nk} and Quinn and Wald \cite{Quinn:1996am}. In the Detweiler--Whiting formulation \cite{Detweiler:2002mi}, the retarded trace-reversed metric perturbation is decomposed as
\begin{equation}
\bar h_{\mu\nu}^{\mathrm{ret}}
=
\bar h_{\mu\nu}^{\mathrm R}
+
\bar h_{\mu\nu}^{\mathrm S},
\label{eq:DW_decomposition}
\end{equation}
where $\bar h_{\mu\nu}^{\mathrm S}$ and $\bar h_{\mu\nu}^{\mathrm R}$ denote the singular and regular fields, respectively.

For convenience, we define the Lorenz-gauge linearized Einstein operator by
\begin{equation}
\mathcal E_{\mu\nu}[\bar h]
\equiv
\Box\bar h_{\mu\nu}
+
2R_{\mu\rho\nu\sigma}\bar h^{\rho\sigma}.
\end{equation}
The retarded and singular fields satisfy the same sourced field equation,
\begin{equation}
\mathcal E_{\mu\nu}
\!\left[
\bar h^{\mathrm{ret}}
\right]
=
-16\pi T_{\mu\nu},
\qquad
\mathcal E_{\mu\nu}
\!\left[
\bar h^{\mathrm S}
\right]
=
-16\pi T_{\mu\nu},
\end{equation}
whereas the regular field satisfies the homogeneous equation
\begin{equation}
\mathcal E_{\mu\nu}
\!\left[
\bar h^{\mathrm R}
\right]
=
0.
\end{equation}

The first-order gravitational self-force is determined by the regular field and its first derivative evaluated on the particle worldline,
\begin{equation}
F^\mu_{\mathrm{self}}(x_p)
=
\mu
k^{\mu\nu\gamma\delta}
\nabla_\delta
\bar h_{\nu\gamma}^{\mathrm R}(x_p),
\label{eq:GSF_regular_field}
\end{equation}
where $x_p$ denotes an event on the worldline and
\begin{equation}
\begin{split}
k^{\mu\nu\gamma\delta}
={}&
\frac{1}{2}g^{\mu\delta}u^\nu u^\gamma
-
g^{\mu\nu}u^\gamma u^\delta
-
\frac{1}{2}u^\mu u^\nu u^\gamma u^\delta
\\
&+
\frac{1}{4}u^\mu g^{\nu\gamma}u^\delta
+
\frac{1}{4}g^{\mu\delta}g^{\nu\gamma}.
\end{split}
\label{eq:force_projection_tensor}
\end{equation}
The structure of $k^{\mu\nu\gamma\delta}$ ensures that the resulting force is orthogonal to the particle four-velocity.

The Detweiler--Whiting singular field is defined only within a normal neighborhood of the worldline and is generally unavailable in a global closed form. 
In practical effective-source calculations, it is replaced by a puncture field $\bar h_{\mu\nu}^{\mathcal P}$, obtained by truncating a local expansion of $\bar h_{\mu\nu}^{\mathrm S}$ at finite order \cite{Barack:2001gx,Heffernan:2012su}. The puncture order is chosen such that
\begin{equation}
\lim_{x\rightarrow x_p}
\left[
\bar h_{\mu\nu}^{\mathcal P}(x)
-
\bar h_{\mu\nu}^{\mathrm S}(x)
\right]
=
0,
\end{equation}
and
\begin{equation}
\lim_{x\rightarrow x_p}
\left[
\nabla_\alpha\bar h_{\mu\nu}^{\mathcal P}(x)
-
\nabla_\alpha\bar h_{\mu\nu}^{\mathrm S}(x)
\right]
=
0.
\end{equation}
We then define the residual field by
\begin{equation}
\label{eq:residual_definition}
\bar h_{\mu\nu}^{\mathcal R}
\equiv
\bar h_{\mu\nu}^{\mathrm{ret}}
-
\bar h_{\mu\nu}^{\mathcal P}.
\end{equation}
We use the superscripts $\mathrm R$ and $\mathcal R$ to distinguish the exact Detweiler--Whiting regular field from the residual field obtained using a finite-order puncture. 
By construction, the two fields and their first derivatives agree on the worldline to the order required for the self-force calculation. 
Equation \eqref{eq:GSF_regular_field} may therefore be evaluated using the residual field:
\begin{equation}
\label{eq:GSF_residual_field}
F^\mu_{\mathrm{self}}(x_p)
=
\mu
k^{\mu\nu\gamma\delta}
\nabla_\delta
\bar h_{\nu\gamma}^{\mathcal R}(x_p).
\end{equation}

Substituting Eq. \eqref{eq:residual_definition} into the linearized Einstein equation gives
\begin{equation}
\label{eq:covariant_effective_source}
\mathcal E_{\mu\nu}
\!\left[
\bar h^{\mathcal R}
\right]
=
-16\pi T_{\mu\nu}
-
\mathcal E_{\mu\nu}
\!\left[
\bar h^{\mathcal P}
\right]
\equiv
S_{\mu\nu}^{\mathrm{eff}}.
\end{equation}
The distributional contribution from the point-particle stress-energy tensor is cancelled by the corresponding distributional part of the puncture-field operator.
The residual field consequently satisfies a wave equation with an effective source that is finite at the particle and whose differentiability is determined by the puncture order.

\subsection{$1+1$D effective source}
\label{sec:gravitational_puncture}
A central ingredient of the present calculation is the analytic construction of the Lorenz-gauge gravitational puncture and effective source for generic Schwarzschild geodesics developed in Ref. \cite{Zhang:2025eqz}. 
Combining a local expansion of the Detweiler--Whiting singular field with a tetrad decomposition and angular modulation, this construction provides all ten tensor-harmonic components required here.
We employ the resulting second-order-in-distance puncture, which yields effective sources that are finite and continuous,
but generally not differentiable, at the particle worldline in the present $1+1$-dimensional formulation.

We use the tensor-harmonic puncture modes
$\bar h_{\ell m}^{(i)\mathcal P}(t,r)$
and their corresponding effective-source modes from Ref. \cite{Zhang:2025eqz}.
The associated numerical implementation is publicly available in Ref. \cite{EffectiveSourceCode}. At the $(\ell,m)$-mode level, the effective source is
\begin{equation}
\label{eq:effective_source_lm}
S_{\ell m}^{(i)\mathrm{eff}}(t,r)
=
S_{\ell m}^{(i)}(t,r)
-
\Box_{2d}
\bar h_{\ell m}^{(i)\mathcal P}(t,r)
-
\mathcal M^{(i)}_{\ (j)}
\bar h_{\ell m}^{(j)\mathcal P}(t,r),
\end{equation}
where summation over the repeated field index $j$ is understood.
The residual-field equations are therefore
\begin{equation}
\label{eq:residual_field_lm}
\Box_{2d}
\bar h_{\ell m}^{(i)\mathcal R}(t,r)
+
\mathcal M^{(i)}_{\ (j)}
\bar h_{\ell m}^{(j)\mathcal R}(t,r)
=
S_{\ell m}^{(i)\mathrm{eff}}(t,r).
\end{equation}

\begin{figure}[htbp]
\centering
\includegraphics[width=0.95\columnwidth]
{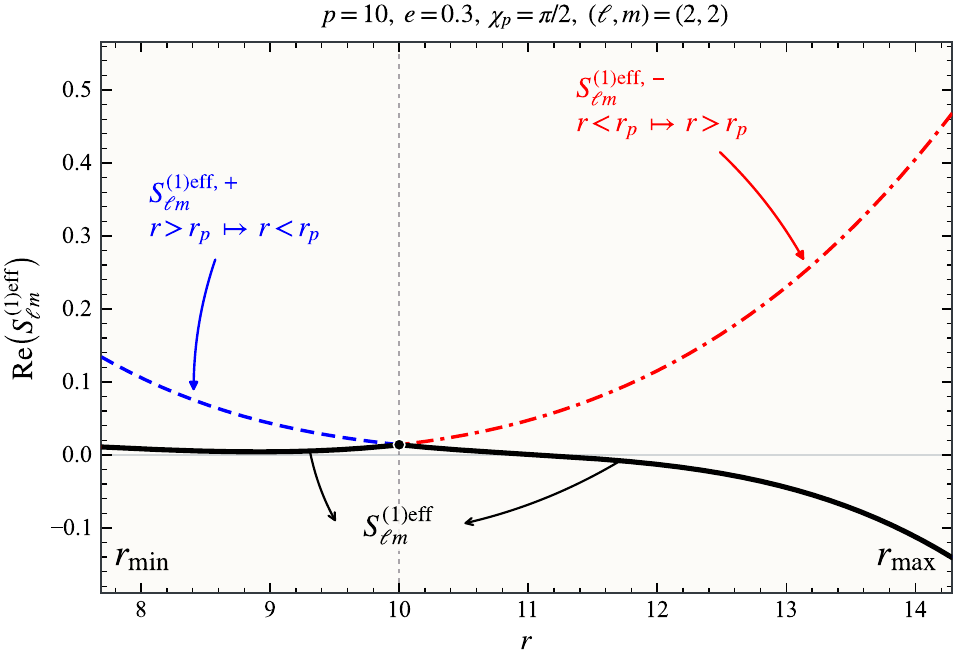}
\caption{
Construction of the analytically extended effective source for the $(\ell,m)=(2,2)$ component of an eccentric orbit with $p=10$ and $e=0.3$, evaluated at $\chi_p=\pi/2$, for which $r_p=10$.
All curves show real parts. 
The black solid line shows the standard (original) effective source $S_{22}^{(1)\mathrm{eff}}$.
The blue dashed line shows $S_{22}^{(1)\mathrm{eff},+}$, obtained by analytically extending the exterior branch of the original effective source from $r>r_p$ into $r<r_p$. 
The red dash-dotted line shows $S_{22}^{(1)\mathrm{eff},-}$, obtained by analytically extending its interior branch from $r<r_p$ into $r>r_p$.
}
\label{fig:extended_efs}
\end{figure}

For the puncture order employed in this work, the effective source is $C^0$ at the worldline. 
As an example, Fig. \ref{fig:extended_efs} shows the time-domain $i=1$ tensor-harmonic component of the $(\ell,m)=(2,2)$ effective source for an eccentric orbit with $(p,e)=(10,0.3)$.
The effective source is evaluated at $\chi_p=\pi/2$, where the particle is located at $r_p=10$.
The solid black curve represents the real part of $S_{22}^{(1)\mathrm{eff}}$.
At this fixed time, the physical effective source is continuous at the particle, but its first radial derivative is discontinuous there.
To compute the self-force in the frequency domain, the time-domain effective source must be Fourier transformed. 
For eccentric motion, a fixed radius within the libration region is crossed twice during each radial period, and the physical source switches between its interior and exterior representations at the corresponding crossing times. 
As a function of coordinate time, it is therefore only finitely differentiable, and its Fourier coefficients decay algebraically with the radial-harmonic index $|n|$.

\subsection{Fourier transformation and Chebyshev representation}
\label{sec:fourier_chebyshev}

Before solving the residual-field Eq. \eqref{eq:residual_field_lm}, we perform a preprocessing step to construct continuous radial representations of the frequency-domain effective source and puncture.
For each tensor-harmonic component, their Fourier modes are first evaluated at a finite set of radial collocation points.
The resulting radial data are then converted into Chebyshev coefficients and stored.
During the subsequent integration of the radial equations, the source and puncture can be evaluated at any required radius directly from their Chebyshev expansions, without repeating the Fourier transformation.
Their Fourier modes are defined by
\begin{equation}
\begin{split}
S_{\ell mn}^{(i){\rm eff}}(r)
&=
\frac{1}{T_r}\int_0^{T_r}S_{\ell m}^{(i){\rm eff}}(t,r)e^{i\omega_{mn}t}\,dt,
\\
\bar h_{\ell mn}^{(i)\mathcal P}(r)
&=
\frac{1}{T_r}\int_0^{T_r}\bar h_{\ell m}^{(i)\mathcal P}(t,r)e^{i\omega_{mn}t}\,dt.
\end{split}
\end{equation}
Equivalently, the time integrals may be evaluated using the relativistic anomaly:
\begin{equation}
\begin{split}
\frac{1}{T_r}\int_0^{T_r}
g(t,r)e^{i\omega_{mn}t}\,dt
&=\frac{1}{T_r}\int_0^{2\pi}
\frac{dt_p}{d\chi}\\
&\times
g\!\bigl[t_p(\chi),r\bigr]
e^{i\omega_{mn}t_p(\chi)}\,d\chi .
\end{split}
\end{equation}

The radial integrations require these frequency-domain modes to be evaluated repeatedly at arbitrary radii. 
Performing the Fourier transformation at every requested integration point would be inefficient.
We therefore evaluate the Fourier modes only at a set of Chebyshev--Lobatto nodes and store their Chebyshev coefficients. 
The source and puncture modes can then be evaluated throughout the libration region directly from their polynomial representations, without repeating the Fourier transformation.

We map the libration region
$r\in[r_{\min},r_{\max}]$ to $x\in[-1,1]$ according to
\begin{equation}
x(r) = \frac{2r-(r_{\max}+r_{\min})}{r_{\max}-r_{\min}},
\end{equation}
with inverse
\begin{equation}
r(x) = \frac{r_{\min}+r_{\max}}{2} + \frac{r_{\max}-r_{\min}}{2}x.
\end{equation}
For $N_{\rm C}$ interpolation points, the Chebyshev--Lobatto nodes are
\begin{equation}
x_q = \cos\left(\frac{\pi q}{N_{\rm C}-1}\right),
\quad
q=0,\ldots,N_{\rm C}-1,
\end{equation}
and the corresponding radial nodes are
\begin{equation}
r_q = \frac{r_{\min}+r_{\max}}{2} + \frac{r_{\max}-r_{\min}}{2}x_q.
\end{equation}

Let $g_q=g(r_q)$ denote the sampled values of a generally complex radial function.
Its Chebyshev interpolant is
\begin{equation}
g(r) \simeq \frac{1}{2}c_0 + \sum_{j=1}^{N_{\rm C}-2} c_jT_j\!\left[x(r)\right] + \frac{1}{2}c_{N_{\rm C}-1} T_{N_{\rm C}-1}\!\left[x(r)\right],
\end{equation}
where
\begin{equation}
T_j(x)=\cos\!\left(j\arccos x\right)
\end{equation}
is the Chebyshev polynomial of the first kind. The coefficients are
\begin{equation}
\begin{split}
c_j=\frac{2}{N_{\rm C}-1}\biggl[
&\frac{g_0+(-1)^j g_{N_{\rm C}-1}}{2}\\
&+\sum_{q=1}^{N_{\rm C}-2}g_q
 \cos\!\left(\frac{\pi jq}{N_{\rm C}-1}\right)
\biggr].
\end{split}
\end{equation}
This is a discrete cosine transform of type I.

For example, the effective-source modes are represented as
\begin{equation}
\begin{split}
S_{\ell mn}^{(i){\rm eff}}(r)\simeq {}
\frac{1}{2}c_{\ell mn,0}^{(i)}
&+\sum_{j=1}^{N_{\rm C}-2}
c_{\ell mn,j}^{(i)}T_j\!\bigl[x(r)\bigr]\\
&+\frac{1}{2}c_{\ell mn,N_{\rm C}-1}^{(i)}
T_{N_{\rm C}-1}\!\bigl[x(r)\bigr],
\end{split}
\end{equation}
where $c_{\ell mn,j}^{(i)}$ denotes the Chebyshev coefficient of the $(\ell,m,n,i)$ effective-source mode.
An independent set of coefficients is constructed for each puncture mode $\bar h_{\ell mn}^{(i)\mathcal P}$.

Because the physical source and puncture are only finitely differentiable, their radial Chebyshev coefficients decay algebraically. 
In the EES formulation developed here, the same procedure is applied separately to the smooth extended fields $S_{\ell mn}^{(i){\rm eff},\pm}$ and $\bar h_{\ell mn}^{(i)\mathcal P,\pm}$,
whose Chebyshev representations converge substantially more rapidly.

\section{Extended Effective-Source Method for Lorenz-Gauge Gravitational Perturbations}
\label{sec:EES}

In this section, we extend the EES method developed for the scalar-field problem in Ref. \cite{Leather:2023dzj} to the coupled Lorenz-gauge gravitational perturbation equations.
The generic-orbit gravitational puncture fields and effective sources described in Sec. \ref{sec:gravitational_puncture} provide the required local regularization input.

\subsection{Extended effective source and puncture}

As discussed above, the effective source (hereafter the standard effective source) is only $C^0$ at the particle worldline.
During eccentric motion, the switching between its branches as the particle moves leads to algebraically decaying Fourier coefficients and hence slow convergence. 
We therefore employ the EES method to accelerate Fourier-series convergence by transforming smooth extensions of the two branches separately.
Within the libration region, we analytically continue the exterior and interior expressions of the standard effective source through the worldline. This defines two smooth functions,
$S_{\ell m}^{(i){\rm eff},+}(t,r)$ and
$S_{\ell m}^{(i){\rm eff},-}(t,r)$, such that
\begin{equation}
\label{eq:extended_source_decomposition}
\begin{split}
S_{\ell m}^{(i){\rm eff}}(t,r)
={}&S_{\ell m}^{(i){\rm eff},+}(t,r)\Theta^+(t,r)\\
&+S_{\ell m}^{(i){\rm eff},-}(t,r)\Theta^-(t,r),
\end{split}
\end{equation}
where
\begin{equation}
\Theta^\pm(t,r) \equiv \Theta\!\left[\pm\bigl(r-r_p(t)\bigr)\right].
\label{eq:heaviside_definition}
\end{equation}
The $+$ extension agrees with the physical exterior expression for $r>r_p(t)$ and continues it into $r<r_p(t)$. 
Similarly, the $-$ extension agrees with the physical interior expression for $r<r_p(t)$ and continues it into $r>r_p(t)$.
Figure \ref{fig:extended_efs} illustrates this construction using the $i=1$ tensor-harmonic component of the time-domain $(\ell,m)=(2,2)$ effective source.
At the selected orbital phase, $\chi_p=\pi/2$, the particle is located at $r_p=10$. 
The physical source coincides with the $-$ branch for $r<r_p$ and with the $+$ branch for $r>r_p$.
Each branch is then smoothly extended across $r=r_p$ into the opposite side of the libration region.

The corresponding frequency-domain modes are
\begin{equation}
S_{\ell mn}^{(i){\rm eff},\pm}(r)
=
\frac{1}{T_r}
\int_0^{T_r}
S_{\ell m}^{(i){\rm eff},\pm}(t,r)
e^{i\omega_{mn}t}\,dt,
\end{equation}
with inverse transformation
\begin{equation}
S_{\ell m}^{(i){\rm eff},\pm}(t,r)
=
\sum_{n=-\infty}^{\infty}
S_{\ell mn}^{(i){\rm eff},\pm}(r)
e^{-i\omega_{mn}t}.
\label{eq:extended_source_inverse}
\end{equation}

The puncture fields are extended in the same manner:
\begin{equation}
\begin{split}
\bar h_{\ell m}^{(i)\mathcal P}(t,r)
={}&\bar h_{\ell m}^{(i)\mathcal P,+}(t,r)\Theta^+(t,r)\\
&+\bar h_{\ell m}^{(i)\mathcal P,-}(t,r)\Theta^-(t,r),
\end{split}
\end{equation}
and 
\begin{equation}
\begin{split}
\bar h_{\ell mn}^{(i)\mathcal P,\pm}(r)
=
\frac{1}{T_r}\int_0^{T_r}\bar h_{\ell m}^{(i)\mathcal P,\pm}(t,r)e^{i\omega_{mn}t}\,dt,
\\
\bar h_{\ell m}^{(i)\mathcal P,\pm}(t,r)
=
\sum_{n=-\infty}^{\infty}\bar h_{\ell mn}^{(i)\mathcal P,\pm}(r)e^{-i\omega_{mn}t}.
\end{split}
\end{equation}

\begin{figure}[htbp]
\centering
\includegraphics[width=0.9\columnwidth]{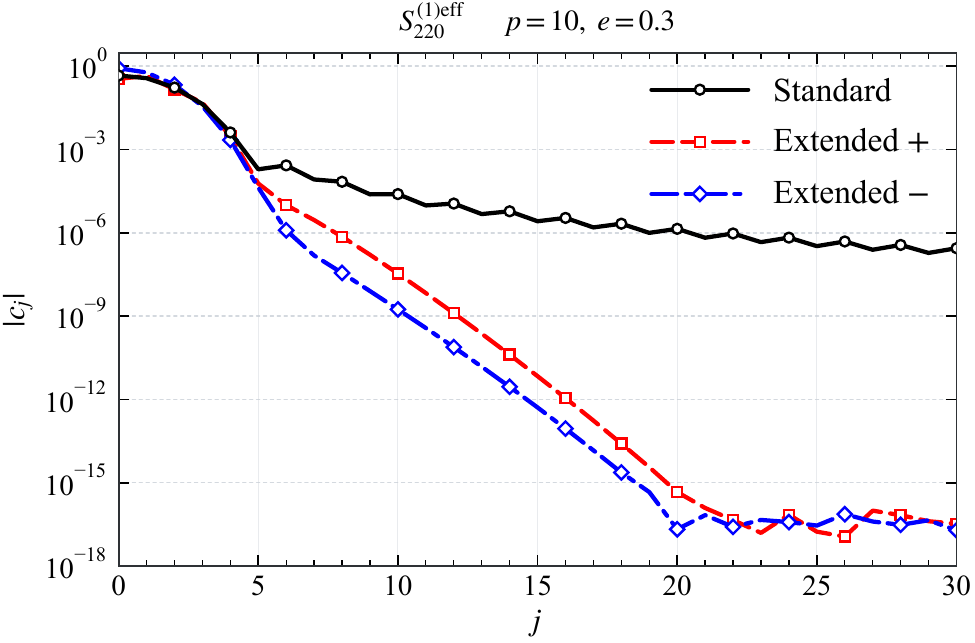}
\caption{Chebyshev coefficient spectra for the $n=0$ Fourier mode of the $i=1$ tensor-harmonic effective source with $(\ell,m)=(2,2)$ and $(p,e)=(10,0.3)$.
The black solid line with circle markers represents the standard effective source; 
the red dashed line with square markers and the blue dashed line with diamond markers represent results of the analytically extended exterior ($+$) and interior ($-$) sources, respectively. 
Both extended-source spectra decay much more rapidly than the standard-source spectrum. 
At $j=30$, the standard-source coefficients remain of order $10^{-6}$--$10^{-7}$, whereas the extended-source coefficients reach the $10^{-16}$--$10^{-17}$ roundoff floor by $j\simeq20$.
Beyond this point, finite floating-point precision sets the numerical floor and prevents the computed coefficients from decreasing further.}
\label{fig:compare_chebyshev}
\end{figure}

At fixed radius, the extended effective sources and puncture fields are smooth functions of coordinate time. Their Fourier coefficients therefore decay rapidly with $|n|$. Each extended frequency-domain mode is represented radially using the Chebyshev interpolation described in Sec. \ref{sec:fourier_chebyshev}.
For the $n=0$ Fourier mode of the $i=1$ tensor-harmonic component with $(\ell,m)=(2,2)$,
Fig. \ref{fig:compare_chebyshev} compares the Chebyshev coefficient spectra of the standard effective source and its two analytic extensions.
The standard effective source retains the branch switching associated with the particle's eccentric motion, and its coefficients exhibit a slowly decaying tail.
By extending the interior and exterior branches separately, we obtain smoother radial functions whose Chebyshev coefficients decay much more rapidly. 
At $j\simeq20$, the standard-source coefficients remain of order $10^{-6}$, while those of the extended sources have fallen to approximately $10^{-17}$, a difference of about ten orders of magnitude. 
The extended-source spectra then reach a floor set by finite floating-point precision. 
This rapid decay allows the extended modes to be represented with fewer Chebyshev coefficients and reduces interpolation error when evaluating the variation-of-parameters integrals for eccentric orbits.
The extended sources can therefore be represented to high accuracy with substantially fewer Chebyshev coefficients, reducing interpolation error in the variation-of-parameters integrals.
For eccentric orbits, where the particle repeatedly crosses radii within the libration region, treating the two smooth branches separately also avoids the slow time-domain Fourier convergence caused by branch switching.

\subsection{Recovery of the frequency-domain modes}
\label{sec:supp_physical_modes}

The unextended frequency-domain modes are required to determine the radiative amplitudes used in the boundary matching. To recover them, we expand the Heaviside functions as
\begin{equation}
\Theta^\pm(t,r)
=
\sum_{s=-\infty}^{\infty}
\beta_s^\pm(r)e^{-is\Omega_rt}.
\label{eq:heaviside_fourier}
\end{equation}

For $r_{\min}<r<r_{\max}$, let $t_\times(r)$ denote the coordinate time elapsed from periapsis to the first outward crossing of radius $r$, with
\begin{equation}
0\leq t_\times(r)\leq\frac{T_r}{2}.
\label{eq:supp_crossing_time}
\end{equation}
The zeroth-order coefficients are
\begin{equation}
\beta_0^+(r)=\frac{2t_\times(r)}{T_r},
\quad
\beta_0^-(r)=1-\frac{2t_\times(r)}{T_r},
\label{eq:supp_beta_zero}
\end{equation}
while for $s\neq0$,
\begin{equation}
\beta_s^\pm(r)
=\pm\frac{1}{\pi s}
\sin\left[
\frac{2\pi s\,t_\times(r)}{T_r}
\right].
\label{eq:supp_beta_nonzero}
\end{equation}
These coefficients satisfy
\begin{equation}
\beta_s^+(r)+\beta_s^-(r)=\delta_{s0}.
\label{eq:supp_beta_identity}
\end{equation}

Substitution of Eqs. \eqref{eq:extended_source_inverse} and \eqref{eq:heaviside_fourier} into Eq. \eqref{eq:extended_source_decomposition} gives
\begin{equation}
\begin{split}
S_{\ell mn}^{(i){\rm eff}}(r)
={}&\sum_{n'=-\infty}^{\infty}
\beta_{n-n'}^+(r)
S_{\ell mn'}^{(i){\rm eff},+}(r)\\
&+\sum_{n'=-\infty}^{\infty}
\beta_{n-n'}^-(r)
S_{\ell mn'}^{(i){\rm eff},-}(r).
\end{split}
\label{eq:physical_source_convolution}
\end{equation}
Similarly, the physical puncture modes are
\begin{equation}
\begin{split}
\bar h_{\ell mn}^{(i)\mathcal P}(r)
={}&\sum_{n'=-\infty}^{\infty}
\beta_{n-n'}^+(r)
\bar h_{\ell mn'}^{(i)\mathcal P,+}(r)\\
&+\sum_{n'=-\infty}^{\infty}
\beta_{n-n'}^-(r)
\bar h_{\ell mn'}^{(i)\mathcal P,-}(r).
\end{split}
\label{eq:physical_puncture_convolution}
\end{equation}

Because $\beta_s^\pm=O(|s|^{-1})$, these convolutions restore the algebraic large-$|n|$ behavior of the physical modes. We use Eqs. \eqref{eq:physical_source_convolution} and \eqref{eq:physical_puncture_convolution} only to determine the physical radiative amplitudes required for matching. The final time-domain residual field is reconstructed directly from the smooth extended solutions, without Fourier reconstructing the Heaviside functions.

\subsection{Extended residual fields}
\label{sec:supp_extended_matching}
We now describe how the residual field is constructed using the extended effective sources and puncture fields.
For a fixed $(\ell,m,n)$ mode, we suppress these indices.
Let $k$ denote the number of independently integrated radial fields.
We construct $k$ linearly independent horizon-ingoing solutions $\bar h_j^{(i)h}(r)$ and $k$ linearly independent infinity-outgoing solutions $\bar h_j^{(i)\infty}(r)$, where $i$ labels a field component and $j=1,\ldots,k$ labels an independent homogeneous solution.

We assemble these solutions into the matrices
\begin{equation}
\label{eq:homogeneous_matrices}
\bm H^h(r)=\left(\bar h_j^{(i)h}(r)\right),
\quad
\bm H^\infty(r)=\left(\bar h_j^{(i)\infty}(r)\right),
\end{equation}
and define the fundamental matrix
\begin{equation}
\label{eq:fundamental_matrix_standard}
\bm\Phi(r)=
\begin{pmatrix}
-\bm H^h(r) & \bm H^\infty(r)
\\
-\partial_{r_*}\bm H^h(r)
&
\partial_{r_*}\bm H^\infty(r)
\end{pmatrix}.
\end{equation}

Let $\bm S^{{\rm eff},\pm}(r)$ denote the $k$-component extended effective-source vectors. The variation-of-parameters integrands are defined by
\begin{equation}
\begin{pmatrix}
\bm v^{h,\pm}(r)
\\
\bm v^{\infty,\pm}(r)
\end{pmatrix}
=
\bm\Phi^{-1}(r)
\begin{pmatrix}
\bm 0
\\
-4\bm S^{{\rm eff},\pm}(r)
\end{pmatrix}
f^{-1}(r).
\label{eq:extended_vop_integrands}
\end{equation}
The corresponding weighting coefficients are
\begin{subequations}
\label{eq:supp_extended_weights}
\begin{align}
C_j^{h,\pm}(r)
&=
\int_r^{r_{\max}}
v_j^{h,\pm}(r')\,dr',
\\
C_j^{\infty,\pm}(r)
&=
\int_{r_{\min}}^r
v_j^{\infty,\pm}(r')\,dr'.
\end{align}
\end{subequations}
The factor $f^{-1}$ in Eq. \eqref{eq:extended_vop_integrands} converts integration with respect to $r_*$ into integration with respect to $r$.

The extended particular solutions are
\begin{equation}
\begin{split}
\bar h^{(i)\mathrm{inh},\pm}(r)
={}&\sum_{j=1}^{k}
C_j^{\infty,\pm}(r)\bar h_j^{(i)\infty}(r)\\
&+\sum_{j=1}^{k}
C_j^{h,\pm}(r)\bar h_j^{(i)h}(r).
\end{split}
\end{equation}
For compactness, we define their homogeneous completions within the libration region by
\begin{equation}
\begin{split}
\mathcal H_\pm^{(i)}(r)
\equiv{}&\bar h^{(i)\mathrm{inh},\pm}(r)\\
&+\sum_{j=1}^{k}
b_j^{\infty,\pm}\bar h_j^{(i)\infty}(r)+\sum_{j=1}^{k}
b_j^{h,\pm}\bar h_j^{(i)h}(r).
\end{split}
\end{equation}

The two extended residual fields then take the form
\begin{equation}
\label{eq:hres}
\begin{split}
\bar h^{(i)\mathcal R,-}(r)
&=
\begin{cases}
\displaystyle
\sum_{j=1}^{k}
a_j^h\bar h_j^{(i)h}(r),
&
r\leq r_{\min},
\\[2mm]
\mathcal H_-^{(i)}(r),
&
r_{\min}<r< r_{\max},
\end{cases}
\\
\bar h^{(i)\mathcal R,+}(r)
&=
\begin{cases}
\mathcal H_+^{(i)}(r),
&
r_{\min}< r<r_{\max},
\\[2mm]
\displaystyle
\sum_{j=1}^{k}
a_j^\infty
\bar h_j^{(i)\infty}(r),
&
r\geq r_{\max}.
\end{cases}
\end{split}
\end{equation}
Here, $a_j^h$ and $a_j^\infty$ are the physical horizon and infinity radiative amplitudes. The coefficients
$b_j^{h,\pm}$ and $b_j^{\infty,\pm}$
supply the homogeneous corrections within the libration region.

To determine these coefficients, we introduce the puncture state vectors
\begin{equation}
\bm\Psi^{\mathcal P,\pm}(r)
=
\begin{pmatrix}
\bar{\bm h}^{\mathcal P,\pm}(r)
\\
\partial_{r_*}\bar{\bm h}^{\mathcal P,\pm}(r)
\end{pmatrix}.
\label{eq:supp_extended_puncture_state}
\end{equation}
At the boundaries of the libration region,
\begin{equation}
\begin{split}
    &\bm\Psi^{\mathcal P,+}(r_{\max})
=\bm\Psi^{\mathcal P}(r_{\max}),\\
&\bm\Psi^{\mathcal P,-}(r_{\min})
=\bm\Psi^{\mathcal P}(r_{\min}).
\end{split}
\end{equation}

The physical amplitudes $a_j^h$ and $a_j^\infty$ are obtained by applying the same variation-of-parameters construction to the physical source modes recovered from Eq. \eqref{eq:physical_source_convolution}. If $C_j^h$ and $C_j^\infty$ denote the corresponding physical weighting coefficients, then
\begin{equation}
\begin{split}
a_j^\infty
={}C_j^\infty(r_{\max})
&-\bigl[
\bm\Phi^{-1}(r_{\min})
\bm\Psi^{\mathcal P}(r_{\min})
\bigr]^\infty_j\\
&+\bigl[
\bm\Phi^{-1}(r_{\max})
\bm\Psi^{\mathcal P}(r_{\max})
\bigr]^\infty_j,\\
a_j^h
={}C_j^h(r_{\min})
&+\bigl[
\bm\Phi^{-1}(r_{\max})
\bm\Psi^{\mathcal P}(r_{\max})
\bigr]^h_j\\
&-\bigl[
\bm\Phi^{-1}(r_{\min})
\bm\Psi^{\mathcal P}(r_{\min})
\bigr]^h_j.
\end{split}
\end{equation}
The superscripts $h$ and $\infty$ outside the square brackets select the blocks associated with the horizon and infinity homogeneous bases, respectively.

Matching the $+$ extended field to the physical infinity-outgoing solution at $r_{\max}$ gives
\begin{equation}
\begin{split}
b_j^{h,+}
&=
\left[
\bm\Phi^{-1}(r_{\max})
\bm\Psi^{\mathcal P,+}(r_{\max})
\right]^h_j,
\\
b_j^{\infty,+}
&=
a_j^\infty
-
C_j^{\infty,+}(r_{\max})\\
&-\left[
\bm\Phi^{-1}(r_{\max})
\bm\Psi^{\mathcal P,+}(r_{\max})
\right]^\infty_j.
\end{split}
\end{equation}
Similarly, matching the $-$ extended field to the physical horizon-ingoing solution at $r_{\min}$ gives
\begin{equation}
\begin{split}
b_j^{\infty,-}
&=
-\left[\bm\Phi^{-1}(r_{\min})
\bm\Psi^{\mathcal P,-}(r_{\min})
\right]^\infty_j,
\\
b_j^{h,-}
&=
a_j^h
-C_j^{h,-}(r_{\min})\\
&+\left[
\bm\Phi^{-1}(r_{\min})
\bm\Psi^{\mathcal P,-}(r_{\min})
\right]^h_j.
\end{split}
\end{equation}
These relations remove the unphysical homogeneous contributions introduced by extending the puncture while preserving the physical retarded amplitudes at the horizon and infinity.

\subsection{Residual field reconstruction}
\label{sec:supp_residual_reconstruction}

The two time-domain extended residual fields are reconstructed independently:
\begin{equation}
\bar h_{\ell m}^{(i)\mathcal R,\pm}(t,r)
=
\sum_{n=-\infty}^{\infty}
\bar h_{\ell mn}^{(i)\mathcal R,\pm}(r)
e^{-i\omega_{mn}t}.
\label{eq:supp_extended_residual_sum}
\end{equation}
The physical residual field is then obtained by applying the Heaviside functions analytically:
\begin{equation}
\label{eq:physical_residual}
\begin{split}
\bar h_{\ell m}^{(i)\mathcal R}(t,r)
={}
&\bar h_{\ell m}^{(i)\mathcal R,+}(t,r)
\Theta^+(t,r)\\
&+
\bar h_{\ell m}^{(i)\mathcal R,-}(t,r)
\Theta^-(t,r).
\end{split}
\end{equation}
Thus, the radial-harmonic sums are performed separately on the two smooth extended fields, and the physical branch switching is imposed only afterward. The Heaviside functions are never approximated in the final time-domain reconstruction by a truncated Fourier series.

After reconstructing the components in the numerical set, we add the corresponding puncture components to recover the retarded-field modes. The Lorenz-gauge constraints are then applied to reconstruct the components in the algebraic Set, after which their puncture components are subtracted. This completes the reconstruction of all ten tensor-harmonic components of the residual metric perturbation.

Because the two extended residual fields are smooth functions of coordinate time at fixed radius, their Fourier sums converge rapidly. The nonsmooth physical branch switching is imposed analytically through Eq. \eqref{eq:physical_residual}, thereby avoiding the slow convergence and Gibbs-type behavior associated with a direct Fourier reconstruction of the physical source.

\section{Numerical Implementation and Results}
\label{sec.nir}

\subsection{Computational algorithm}

We now summarize the EES algorithm for computing the Lorenz-gauge GSF on an eccentric Schwarzschild orbit.

\begin{enumerate}[label={(\arabic*)}]

\item \emph{Orbital parameters.}
For a specified semilatus rectum $p$ and eccentricity $e$, we compute the radial turning points $r_{\mathrm{min}}$ and $r_{\mathrm{max}}$, the constants of motion $\mathcal E$ and $\mathcal L$, and the orbital quantities $T_r$, $\Omega_r$, and $\Omega_\phi$.

\item \emph{Numerical grid.}
We work in the tortoise coordinate $r_*$. The inner numerical boundary is placed near the event horizon at
$r_{*,\mathrm{in}}=-50,$
while the outer boundary is placed in the wave zone at
$r_{*,\mathrm{out}}=10/|\omega|$.
Within the libration region
$r_*\in[r_{*,\mathrm{min}},r_{*,\mathrm{max}}]$,
we introduce a uniform grid of $N_{\mathrm{grid}}=5000$ points. This grid is used to store the homogeneous basis solutions and facilitate their high-order interpolation during the radial integrations.

\item \emph{Boundary conditions.}
For each dynamical $(\ell,m,n)$ mode, we impose purely ingoing boundary conditions at the event horizon and purely outgoing boundary conditions at spatial infinity.
The initial field values and their radial derivatives are obtained by evaluating the near-horizon Frobenius expansions and large-radius asymptotic expansions at $r_{*,\mathrm{in}}$ and $r_{*,\mathrm{out}}$, respectively \cite{Akcay:2010dx,Akcay:2013wfa}.
For static modes, the radiative boundary conditions are replaced by regularity at the future event horizon and spatial infinity.

\item \emph{Homogeneous fields.}
For each $(\ell,m,n)$ mode, we solve the coupled radial equations for the $k$ fields in the numerical set using the adaptive Prince--Dormand Runge--Kutta integrator \texttt{rk8pd}.
A complete $2k$-dimensional homogeneous basis is constructed by integrating the infinity-outgoing solutions
$\bar h_j^{(i)\infty}(r)$
inwards from $r_{*,\mathrm{out}}$ to $r_{*,\mathrm{min}}$ and the horizon-ingoing solutions
$\bar h_j^{(i)h}(r)$
outwards from $r_{*,\mathrm{in}}$ to $r_{*,\mathrm{max}}$.
Both bases are consequently available throughout the libration region, allowing the $2k\times2k$ fundamental matrix in Eq. \eqref{eq:fundamental_matrix_standard} to be constructed at every required radial point.

Near-static modes require special care because the fundamental matrix becomes increasingly ill conditioned as $|\omega_{mn}|$ approaches zero. 
Following Ref. \cite{Akcay:2013wfa}, we condition the outer homogeneous basis using the weak-field eigensolutions and rescale the corresponding boundary amplitudes by appropriate powers of $\omega_{mn}$. 
This rescaling balances the magnitudes of the inward-integrated homogeneous solutions in the libration region and improves the conditioning of the variation-of-parameters system. 
The exactly static modes, $\omega_{mn}=0$, are treated separately. 
An alternative treatment based on thin-QR preconditioning was developed in Ref. \cite{Osburn:2014hoa}.

\item \emph{Extended effective sources and puncture fields.}
We evaluate the gravitational puncture fields and effective sources using the analytic generic-orbit construction of Ref. \cite{Zhang:2025eqz}, whose associated numerical implementation is available in Ref. \cite{EffectiveSourceCode}.
We analytically continue the exterior and interior expressions through the worldline to construct
$S_{\ell m}^{(i)\mathrm{eff},\pm}(t,r)$,
and 
$\bar h_{\ell m}^{(i)\mathcal P,\pm}(t,r)$.

Their Fourier transforms yield the extended frequency-domain modes
$S_{\ell mn}^{(i)\mathrm{eff},\pm}(r)$,
and
$\bar h_{\ell mn}^{(i)\mathcal P,\pm}(r)$.
These radial functions are represented using the Chebyshev interpolation procedure described in Sec. \ref{sec:fourier_chebyshev}.

\item \emph{Variation-of-parameters coefficients.}
For each extended source, we compute the weighting coefficients
$C_j^{h,\pm}(r)$
and
$C_j^{\infty,\pm}(r)$
by integrating the source vector against the appropriate blocks of the inverse fundamental matrix.
The physical, unextended source and puncture modes reconstructed through Eqs. \eqref{eq:physical_source_convolution} and \eqref{eq:physical_puncture_convolution} determine the radiative amplitudes $a_j^h$ and $a_j^\infty$.
Matching each extended field to these physical radiative solutions at the appropriate boundary then determines the internal coefficients
$b_j^{h,\pm}$ and $b_j^{\infty,\pm}$.

\item \emph{Frequency-domain reconstruction.}
The weighting and matching coefficients are used to construct the extended residual modes
$\bar h_{\ell mn}^{(i)\mathcal R,\pm}(r)$.
For each extension, we first perform the Fourier sum over $n$ to obtain $\bar h_{\ell m}^{(i)\mathcal R,\pm}(t,r)$.
The physical residual field is then recovered by applying the Heaviside functions analytically according to Eq. \eqref{eq:physical_residual}. 
In this way, the nonsmooth switching between the two physical branches is imposed exactly rather than approximated by a truncated Fourier series.

\item \emph{Algebraic reconstruction.}
After summing over the radial harmonics, we obtain the components in the numerical set of the residual metric perturbation at the particle. 
We then add the corresponding puncture components to recover the retarded field modes and apply the Lorenz-gauge constraints to determine the components in the algebraic set. 
Finally, the corresponding puncture components are subtracted, completing the reconstruction of all ten tensor-harmonic components of the residual metric perturbation for each $(\ell,m)$ mode.

\item \emph{Gravitational self-force.}
Using the reconstructed residual metric perturbation and its derivatives at the particle, we evaluate the gravitational self-force following the procedure described in Sec. \ref{sec:self_force_reconstruction}. 
Since the singular field has already been removed through the puncture, no conventional mode-sum regularization parameters are required. 
In the present calculation, the tensor-harmonic sum is truncated at $\ell_{\max}=15$.

\end{enumerate}

\subsection{Self-force reconstruction from the residual field}
\label{sec:self_force_reconstruction}

We construct the self-force from the residual metric
perturbation following Barack and Sago \cite{Barack:2010tm},
with the same off-worldline extension of the force operator.
Substituting the residual tensor-harmonic amplitudes into
this operator and taking $t\to t_p$ and $r\to r_p^\pm$,
while retaining the angular dependence, gives
\begin{equation}
\begin{split}
\left[F_{\mathcal R}^{\alpha}
(\theta,\phi;t_p,r_p)\right]^\pm
=&\frac{\mu^2}{r_p^2}
\sum_{\ell=0}^{\infty}\sum_{m=-\ell}^{\ell}
\Bigl\{
 f_{0\pm}^{\alpha\ell m}Y^{\ell m}\\
 &+f_{1\pm}^{\alpha\ell m}\sin^2\theta\,Y^{\ell m}\\
&
 +f_{2\pm}^{\alpha\ell m}
  \cos\theta\sin\theta\,Y_{,\theta}^{\ell m}\\
&
 +f_{3\pm}^{\alpha\ell m}
  \sin^2\theta\,Y_{,\theta\theta}^{\ell m}\\
&
 +f_{4\pm}^{\alpha\ell m}
  \bigl(\cos\theta\,Y^{\ell m}
        -\sin\theta\,Y_{,\theta}^{\ell m}\bigr)\\
&
 +f_{5\pm}^{\alpha\ell m}
  \sin\theta\,Y_{,\theta}^{\ell m}\\
&
 +f_{6\pm}^{\alpha\ell m}
  \sin^3\theta\,Y_{,\theta}^{\ell m}\\
&
 +f_{7\pm}^{\alpha\ell m}
  \cos\theta\sin^2\theta\,
  Y_{,\theta\theta}^{\ell m}
\Bigr\}.
\end{split}
\label{eq:residual_force_angular}
\end{equation}
Here $f_{A\pm}^{\alpha\ell m}$ are linear combinations of
$\bar h_{\ell m}^{(i)\mathcal R}$ and their first derivatives,
evaluated at $(t_p,r_p^\pm)$.
They have the same expressions as the retarded-field
coefficients in the cited construction, with the metric
amplitudes and their derivatives replaced by the residual
ones.

The angular factors in
Eq. \eqref{eq:residual_force_angular} couple neighboring
spherical harmonics. Re-expanding them in $Y^{\ell m}$ and
evaluating at $(\theta_p,\phi_p)=(\pi/2,\phi_p)$ yields
\begin{equation}
\begin{split}
\left[F_{\mathcal R}^{\alpha\ell}(x_p)\right]^\pm
=&\frac{\mu^2}{r_p^2}
\sum_{m=-\ell}^{\ell}
Y^{\ell m}(\pi/2,\phi_p)\Bigl[\\
&
\mathcal F_{(-3)}^{\alpha,\ell-3,m}
+\mathcal F_{(-2)}^{\alpha,\ell-2,m}
+\mathcal F_{(-1)}^{\alpha,\ell-1,m}
+\mathcal F_{(0)}^{\alpha\ell m}
\\
&+\mathcal F_{(+1)}^{\alpha,\ell+1,m}
+\mathcal F_{(+2)}^{\alpha,\ell+2,m}
+\mathcal F_{(+3)}^{\alpha,\ell+3,m}
\Bigr]^\pm.
\end{split}
\end{equation}

The quantities $\mathcal F_{(k)}^{\alpha\ell'm}$ are the
linear combinations of $f_{A\pm}^{\alpha\ell'm}$ generated
by the angular re-expansion; the subscript $k$ labels
the contribution from tensor degree $\ell'=\ell+k$
to scalar degree $\ell$.
Terms with $\ell'<0$ or $|m|>\ell'$ vanish.
Thus, a force mode of scalar degree $\ell$ contains
tensor-harmonic contributions with
$\ell-3\leq\ell'\leq\ell+3$.

For the puncture order used here, the two radial limits
agree at the particle in the continuum limit.
Suppressing the $\pm$ label, the self-force is
\begin{equation}
F_{\rm self}^{\alpha}(x_p)
=
\sum_{\ell=0}^{\infty}
F_{\mathcal R}^{\alpha\ell}(x_p).
\label{eq:supp_direct_residual_sum}
\end{equation}
The singular contribution has already been removed by
the puncture, so the leading mode-sum regularization
parameters need not be subtracted.

Numerically, we retain force modes through
$\ell_{\max}=15$ and estimate the remaining contribution
from their large-$\ell$ asymptotic behavior:
\begin{equation}
F_{\rm self}^{\alpha}(x_p)
\simeq
\sum_{\ell=0}^{\ell_{\max}}
F_{\mathcal R}^{\alpha\ell}(x_p)
+
F_{\rm tail}^{\alpha}(x_p),
\label{eq:supp_self_force_tail}
\end{equation}
where $F_{\rm tail}^{\alpha}$ accounts for
$\ell>\ell_{\max}$.

\subsection{Numerical results and validation}
\label{sec:results}

\begin{figure}[H]
\centering
\includegraphics[width=0.80\columnwidth]{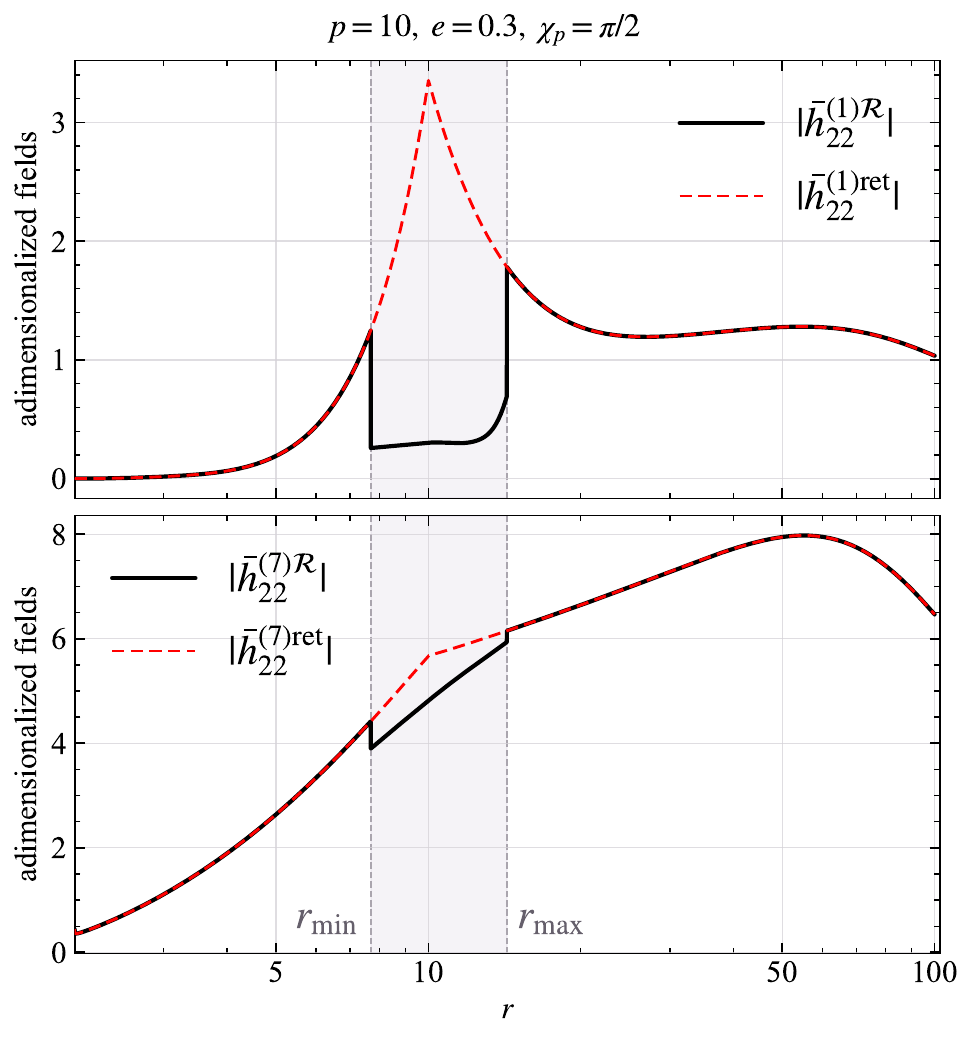}
\caption{Moduli of the retarded and residual $(\ell,m)=(2,2)$ fields for an eccentric orbit with $(p,e)=(10,0.3)$ at $\chi_p=\pi/2$, where the particle is at $r_p=10$.
The upper and lower panels show the $i=1$ and $i=7$ tensor-harmonic components, respectively.
In each panel, the red dashed curve is $|\bar h_{22}^{(i)\mathrm{ret}}|$ and the black solid curve is $|\bar h_{22}^{(i)\mathcal R}|$.
The vertical dashed lines bound the shaded libration region; outside it, the puncture vanishes and the two fields coincide.
At the particle, the retarded field is continuous but has a discontinuous first radial derivative,
whereas puncture subtraction leaves the residual field continuously differentiable.}
\label{fig:residual_main}
\end{figure}

\begin{figure}[H]
\centering
\includegraphics[width=0.75\columnwidth]{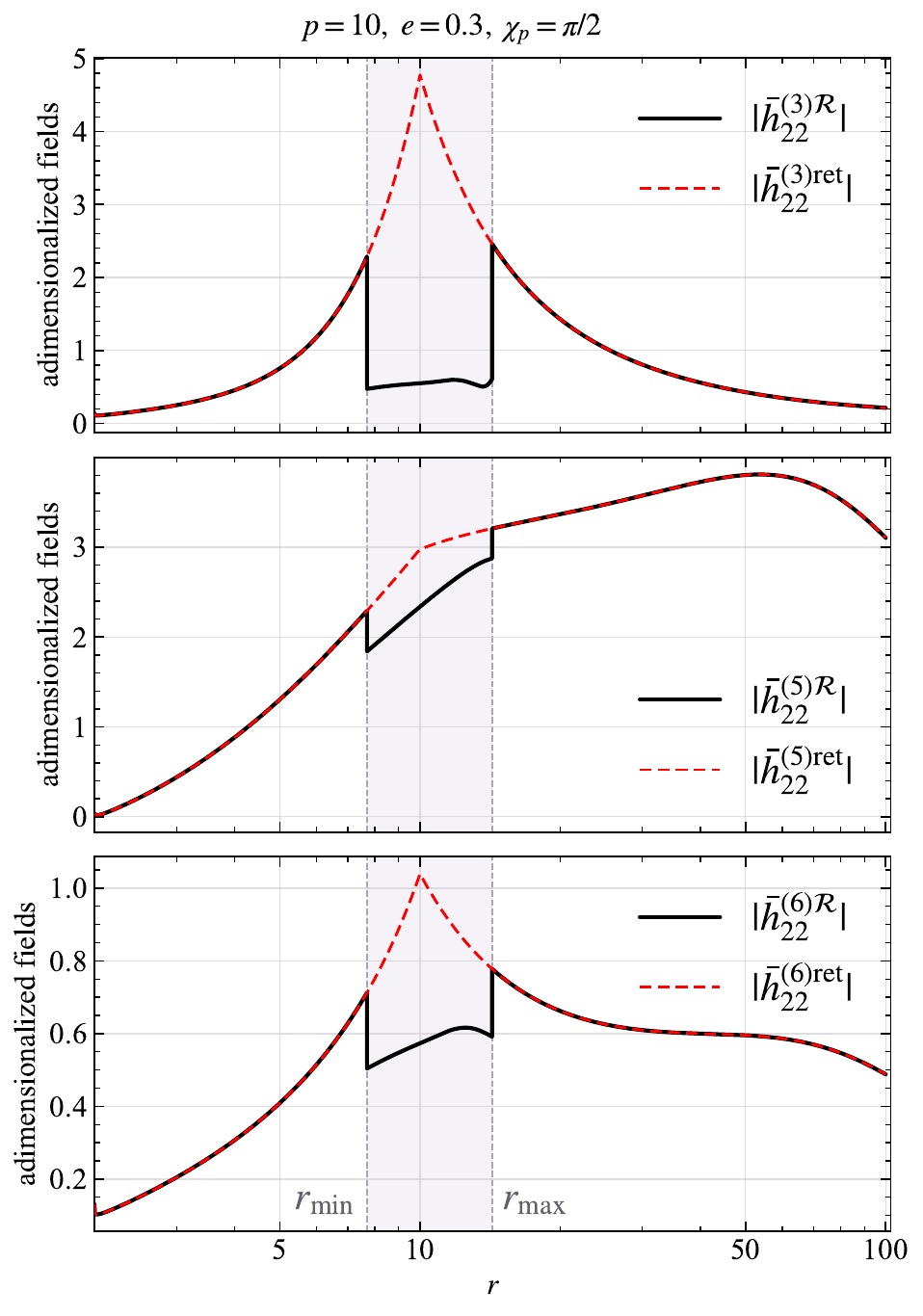}
\caption{Same as Fig. \ref{fig:residual_main} but with $i=3, 5, 6$ tensor-harmonic components.}
\label{fig:residual_h3h5h6}
\end{figure}

We illustrate the residual-field reconstruction for an eccentric orbit with $(p,e)=(10,0.3)$. 
Figures \ref{fig:residual_main} and \ref{fig:residual_h3h5h6} compare the moduli of the retarded and residual fields for the $i=1,3,5,6,7$ tensor-harmonic components of the $(\ell,m)=(2,2)$ mode at $\chi_p=\pi/2$. 
The retarded fields are computed independently using the method of extended homogeneous solutions \cite{Akcay:2013wfa,Osburn:2014hoa}.
Outside the libration region, the puncture vanishes and the residual and retarded fields coincide. 
Within the libration region, puncture subtraction removes the kink at the particle: the retarded fields are continuous but have discontinuous first radial derivatives at $r=r_p$, while the residual fields are continuously differentiable there.
The figures thus illustrate the regularity gained at the particle and the agreement of the two fields outside the puncture-supported region.

\begin{figure}[H]
\centering
\includegraphics[width=0.90\columnwidth]{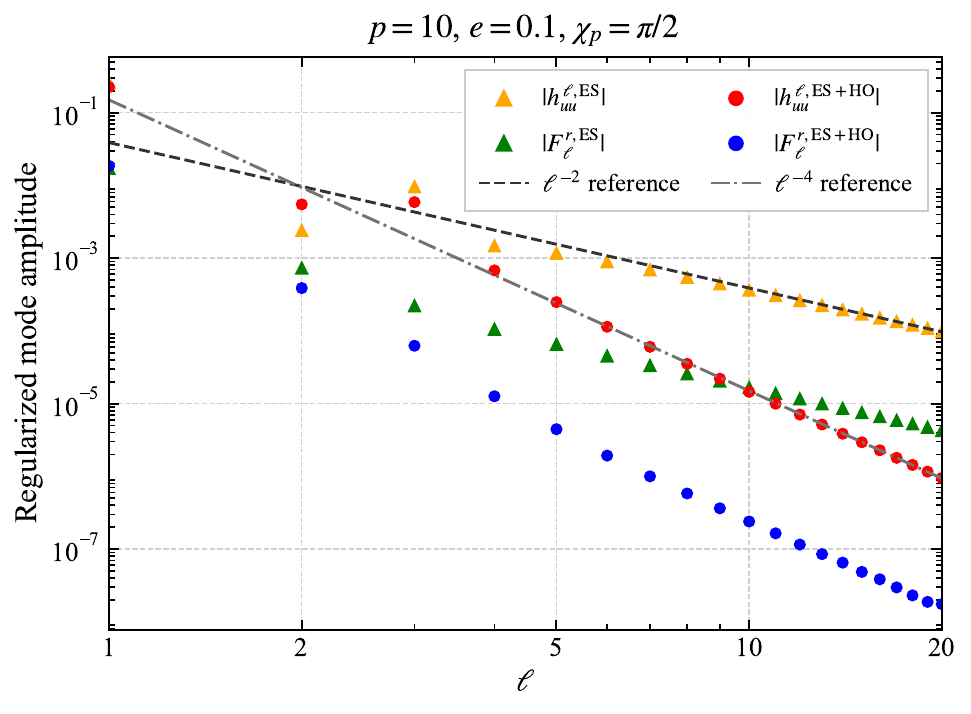}
\caption{Large-$\ell$ behavior of the absolute regularized mode contributions to $h_{uu}^{\ell,\mathcal R}$ and $F^r_\ell$ for $(p,e)=(10,0.1)$ at $\chi_p=\pi/2$.
Triangles show the effective-source (ES) results without higher-order regularization terms: orange for $h_{uu}^{\ell,\mathcal R}$ and green for $F^r_\ell$.
Circles show the results with these terms included (ES+HO): red for $h_{uu}^{\ell,\mathcal R}$ and blue for $F^r_\ell$.
The dashed and dash-dotted lines indicate the $\ell^{-2}$ and $\ell^{-4}$ reference scalings, respectively.
With higher-order regularization terms included, the observed large-$\ell$ decay of both $h_{uu}^{\ell,\mathcal R}$ and $F^r_\ell$ improves from approximately $\ell^{-2}$ to a rate consistent with $\ell^{-4}$.}
\label{fig:eff_convergence}
\end{figure}

Because we use an effective source constructed from a second-order-in-distance puncture field,
the residual modes retain a slowly decaying large-$\ell$ tail.
This limits the accuracy attainable with a truncated mode sum.
Figure \ref{fig:eff_convergence} shows the absolute $\ell$-mode contributions to the radial self-force $F^r$ and to $h_{uu}^{R}\equiv h_{\mu\nu}^{R}u^\mu u^\nu$ \cite{Detweiler:2008ft} for an orbit with $(p,e)=(10,0.1)$ at $\chi_p=\pi/2$.
Before the higher-order regularization correction \cite{Heffernan:2012su}, the modes of both quantities decay approximately as $\ell^{-2}$.
After the correction removes their leading large-$\ell$ contribution, both sets of modes exhibit decay consistent with $\ell^{-4}$.
The $h_{uu}^{R}$ results provide a complementary check of this improvement.
The faster decay reduces the contribution of the unresolved high-$\ell$ tail and improves the accuracy of the self-force at a fixed mode cutoff.

To test whether the improved large-$\ell$ convergence also yields an accurate orbital observable, we compute the first-order GSF contribution to the orbit-averaged generalized redshift. 
We write
$\langle U\rangle=T_r/\mathcal T_r
=\langle U\rangle_0+q\langle U\rangle_{\rm gsf}+O(q^2)$,
where $q=\mu/M$, and $T_r$ and $\mathcal T_r$ are the coordinate-time and proper-time radial periods.
Following Ref. \cite{Akcay:2015pza}, we obtain the correction at fixed orbital frequencies from
\begin{equation}
q\langle U\rangle_{\rm gsf}
=
\frac{1}{2}\langle U\rangle_0
\langle h_{uu}^{R}\rangle_{\tau_0}
+\alpha\mathcal E\langle U\rangle_0^2,
\label{eq:averaged_redshift_gsf}
\end{equation}
where $\langle U\rangle_0=T_{r0}/\mathcal T_{r0}$,
and $\alpha=-\frac{1}{2}\lim_{r\to\infty}h_{tt}^{\rm ret}$ accounts for the asymptotic time-normalization correction.


The proper-time average of the regularized field is evaluated over the full radial period:
\begin{equation}
\langle h_{uu}^{R}\rangle_{\tau_0}=\frac{1}{\mathcal T_{r0}}\int_0^{2\pi}h_{uu}^{R}(\chi)\frac{d\tau_0}{d\chi}\,d\chi.
\label{eq:averaged_huu}
\end{equation}
The metric perturbations in these expressions are given in the original Lorenz gauge.
Table \ref{tab:averaged_redshift} compares our EES results for $(p,e)=(10,0.1)$ and $(10,0.3)$ with those of
Ref. \cite{Akcay:2015pza}.
For $e=0.1$, the results agree to within $10^{-7}$.
For $e=0.3$, the difference is
$1.4\times10^{-6}$, or about $1.2\times10^{-5}$ in relative terms.
The comparison shows that the EES method, combined with higher-order mode-sum regularization and large-$\ell$ tail fitting, can produce accurate orbit-averaged results for eccentric motion.

\begin{table}[H]
\centering
\caption{Comparison of the first-order GSF contribution to the orbit-averaged generalized redshift, $\langle U\rangle_{\rm gsf}$.
The present calculation includes corrected scalar-harmonic force modes through $\ell_{\max}=15$, an estimated large-$\ell$ tail and the asymptotic time-normalization correction.
}
\label{tab:averaged_redshift}
\begin{tabular*}{\linewidth}{@{\extracolsep{\fill}}cccc@{}}
\toprule
$p$ & $e$ & This work & Ref. \cite{Akcay:2015pza} \\
\midrule
$10$ & $0.1$ & $-0.1277540$ & $-0.1277540(3)$ \\
$10$ & $0.3$ & $-0.1168034$ & $-0.1168020(6)$ \\
\bottomrule
\end{tabular*}
\end{table}

Finally, we evaluate the first-order gravitational self-force for three eccentric orbits with $p=10$. Tables \ref{tab:e01}, \ref{tab:e03}, and \ref{tab:e05} compare our dimensionless force components, $\hat F^\alpha=(M/\mu)^2F^\alpha_{\mathrm{self}}$, with the independent results of Ref. \cite{Osburn:2014hoa} for $e=0.1$, $0.3$, and $0.5$, respectively, at five orbital phases.
For each orbit, we apply the higher-order regularization correction, sum the force modes through $\ell_{\max}=15$, and estimate the remaining contribution from $\ell>15$ by fitting the large-$\ell$ behavior of the corrected modes.

The results agree closely with the reference calculation across all three eccentricities. 
Based on the displayed values, the largest relative differences are approximately $9\times10^{-7}$, $1.9\times10^{-5}$, and $4.3\times10^{-4}$ for $e=0.1$, $0.3$, and $0.5$, respectively. 
In particular, the radial-force values for $e=0.3$ agree to all displayed digits at every listed phase. 
Although the differences increase at $e=0.5$, all tabulated components remain within $5\times10^{-4}$ of the reference values in relative terms.
These comparisons show that the gravitational EES calculation, together with higher-order regularization and large-$\ell$ tail fitting, yields accurate self-force values over the range of eccentricities considered here.
Agreement with an independent calculation also provides an end-to-end check of the effective source, field reconstruction, and force evaluation.

\begin{table}[H]
\centering
\caption{Comparison of the dimensionless GSF components
$\hat F^\alpha=(M/\mu)^2F^\alpha_{\rm self}$ for
$(p,e)=(10,0.1)$.
The present calculation includes corrected scalar-harmonic force modes through $\ell_{\max}=15$ and an estimated large-$\ell$ tail.}
\label{tab:e01}
\begin{tabular*}{\linewidth}{@{\extracolsep{\fill}}cccc@{}}
\toprule
Component & $\chi_p$ & This work & Ref. \cite{Osburn:2014hoa} \\
\midrule
$\hat F^t$ & $0$       & $-2.262916\times10^{-4}$ & $-2.262915\times10^{-4}$ \\
            & $\pi/4$   & $ 1.198168\times10^{-4}$ & $ 1.198168\times10^{-4}$ \\
            & $\pi/2$   & $ 2.767753\times10^{-4}$ & $ 2.767753\times10^{-4}$ \\
            & $3\pi/4$  & $ 1.810961\times10^{-4}$ & $ 1.810961\times10^{-4}$ \\
            & $\pi$     & $-3.133613\times10^{-5}$ & $-3.133613\times10^{-5}$ \\
\midrule
$\hat F^r$ & $0$       & $1.606775\times10^{-2}$ & $1.606774\times10^{-2}$ \\
            & $\pi/4$   & $1.544991\times10^{-2}$ & $1.544991\times10^{-2}$ \\
            & $\pi/2$   & $1.360190\times10^{-2}$ & $1.360189\times10^{-2}$ \\
            & $3\pi/4$  & $1.180704\times10^{-2}$ & $1.180704\times10^{-2}$ \\
            & $\pi$     & $1.105403\times10^{-2}$ & $1.105404\times10^{-2}$ \\
\bottomrule
\end{tabular*}
\end{table}

\begin{table}[H]
\centering
\caption{Same as Table \ref{tab:e01} but with $e=0.3$.}
\label{tab:e03}
\begin{tabular*}{\linewidth}{@{\extracolsep{\fill}}cccc@{}}
\toprule
Component & $\chi_p$ & This work & Ref. \cite{Osburn:2014hoa} \\
\midrule
$\hat F^t$&$0$&$-1.02425\times10^{-3}$&$-1.02425\times10^{-3}$\\
&$\pi/4$&$7.93710\times10^{-4}$&$7.93725\times10^{-4}$\\
&$\pi/2$&$1.12071\times10^{-3}$&$1.12072\times10^{-3}$\\
&$3\pi/4$&$5.23325\times10^{-4}$&$5.23325\times10^{-4}$\\
&$\pi$&$2.83617\times10^{-7}$&$2.83617\times10^{-7}$\\
\midrule
$\hat F^r$&$0$&$2.30316\times10^{-2}$&$2.30316\times10^{-2}$\\
&$\pi/4$&$2.10318\times10^{-2}$&$2.10318\times10^{-2}$\\
&$\pi/2$&$1.41875\times10^{-2}$&$1.41875\times10^{-2}$\\
&$3\pi/4$&$8.91644\times10^{-3}$&$8.91644\times10^{-3}$\\
&$\pi$&$7.11090\times10^{-3}$&$7.11090\times10^{-3}$\\
\bottomrule
\end{tabular*}
\end{table}

\begin{table}[H]
\centering
\caption{Same as Table \ref{tab:e01} but with $e=0.5$.}
\label{tab:e05}
\begin{tabular*}{\linewidth}{@{\extracolsep{\fill}}cccc@{}}
\toprule
Component & $\chi_p$ & This work & Ref. \cite{Osburn:2014hoa} \\
\midrule
$\hat F^t$&$0$&$-3.6591\times10^{-3}$&$-3.6577\times10^{-3}$\\
&$\pi/4$&$2.3240\times10^{-4}$&$2.3230\times10^{-4}$\\
&$\pi/2$&$2.1670\times10^{-3}$&$2.1668\times10^{-3}$\\
&$3\pi/4$&$6.5637\times10^{-4}$&$6.5637\times10^{-4}$\\
&$\pi$&$1.0092\times10^{-6}$&$1.0093\times10^{-6}$\\
\midrule
$\hat F^r$&$0$&$3.3854\times10^{-2}$&$3.3855\times10^{-2}$\\
&$\pi/4$&$3.0193\times10^{-2}$&$3.0193\times10^{-2}$\\
&$\pi/2$&$1.5020\times10^{-2}$&$1.5020\times10^{-2}$\\
&$3\pi/4$&$6.3892\times10^{-3}$&$6.3892\times10^{-3}$\\
&$\pi$&$3.9352\times10^{-3}$&$3.9352\times10^{-3}$\\
\bottomrule
\end{tabular*}
\end{table}

\section{Conclusion}
\label{sec:conclusion}

Building on the scalar-field EES method of Ref. \cite{Leather:2023dzj} and the analytic Lorenz-gauge gravitational puncture and effective source of Ref. \cite{Zhang:2025eqz},
we develop a gravitational EES framework for eccentric Schwarzschild orbits and apply it to compute the first-order gravitational self-force.
We analytically extend the interior and exterior expressions of the effective source and puncture before Fourier decomposition. This removes the moving branch switch from the quantities being decomposed and yields rapidly convergent Fourier series. We then solve the coupled frequency-domain equations and reconstruct the physical residual field from the extended solutions.

The numerical results demonstrate the benefit of this construction.
The Chebyshev coefficients of the extended sources reach the floating-point roundoff floor near $j\simeq20$, while those of the standard source remain of order $10^{-6}$. 
Including higher-order regularization terms improves the large-$\ell$ decay of the self-force modes from approximately $\ell^{-2}$ to $\ell^{-4}$.
With an estimate of the remaining large-$\ell$ tail, our self-force results for eccentricities $e=0.1$, $0.3$, and $0.5$ agree closely with independent Lorenz-gauge calculations.
These results demonstrate that gravitational EES can produce accurate self-force values for eccentric orbits.

This first-order calculation provides a proof of concept for gravitational EES on eccentric orbits.
Extending the framework to second order will require a puncture and effective source adapted to eccentric motion \cite{Wardell:2021fyy,Miller:2023ers,Upton:2025bja}; 
treatment of the quadratic source and solution of the second-order perturbation equations, including Teukolsky formulations \cite{Spiers:2023mor,Spiers:2023cip,Leather:2026zhl};
and control of infrared and memory effects at large radius \cite{Cunningham:2024dog,Spiers:2026yqx}.

\begin{acknowledgments}
We are grateful to Sarp Akcay, Benjamin Leather and Thomas Osburn for valuable discussions. 
This research is supported in part by the National Natural Science Foundation of China under Grant Nos. 12535002, 12588101 and 12505076, and the Ningbo Science \& Technology Bureau under program grant No. 2024DJRC010009.

\end{acknowledgments}

\end{CJK*}


%

\end{document}